\documentclass{jltp}

\usepackage{graphicx} 

\title{Vortex Lattice Structures of a Bose-Einstein Condensate in a Rotating Lattice Potential}

\author{Toshihiro Sato$^*$, Tomohiko Ishiyama and Tetsuro Nikuni}

\address{The Institute for Solid State Physics, The University of Tokyo, 5-1-5 Kashiwanoha,\\
                                Kashiwa, Chiba 277-8581, Japan$^*$\\
         Tokyo University of Science, 1-3 Kagurazaka, Shinjuku-ku, Tokyo, 162-9601, Japan}

\runninghead{T. Sato, T. Ishiyama and T. Nikuni}{Vortex Lattice Structures of BEC in a Rotating Lattice Potential}

\begin{document}

\maketitle

\begin{abstract}
We study vortex lattice structures of a trapped Bose-Einstein condensate in 
a rotating lattice potential by numerically solving the time-dependent 
Gross-Pitaevskii equation.
By rotating the lattice potential, we observe the transition from the Abrikosov
vortex lattice to the pinned lattice.
We investigate the transition of the vortex lattice structure by changing conditions such as angular velocity, intensity, and lattice constant of the rotating lattice potential. 

\end{abstract}

Quantized vortices are one of the most characteristic manifestations of superfluidity
associated with a Bose-Einstein condensate (BEC) in atomic gases.
By rotating anisotropic trap potentials, several experimental groups observed
formation of triangular Abrikosov lattices of vortices in rotating BECs \cite{KM,FC,JA}.
Microscopic mechanism of the vortex lattice formation has been extensively studied
both analytically and numerically using the Gross-Pitaevskii (GP) equation
for the condensate wavefunction \cite{DF,MT,MU,KK,NS,KKA,AF,TS}.
More recently, the vortex phase diagrams of a BEC in rotating lattice potentials
have attracted theoretical attention, since one expects vortex pinning and
structural phase transition of vortex lattice structures \cite{RB,JR,HP}.
Recently, a rotating lattice has been experimentally realized at JILA, makinig
use of a laser beam passing through a rotating mask \cite{RB}.
Stimulated by the recent JILA experiment, in this paper, we study vortex lattice
structures of a BEC in a rotating triangular lattice potential created by blue-detuned laser
beams.
%
\begin{figure}
\centerline{\includegraphics[height=1.4in]{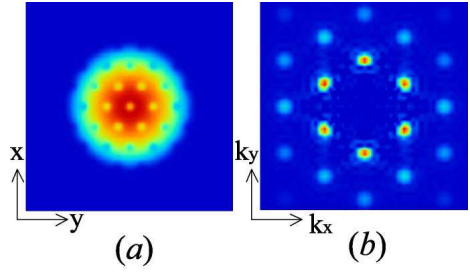}}
\caption{ Density profile $(a)$ and Structure factor profile $(b)$
of BEC in the lattice potential. The lattice potential geometry is triangular lattice
at $a/a_{ho}=2.2$ and $\sigma/a_{ho}=0.65$.}  
\label{fig:tau2}
\end{figure}

We numerically solve the two-dimensional Gross-Pitaevskii (GP) equation in a frame 
rotationg with anguler velocity $\Omega$:
\begin{eqnarray}
(i-\gamma)\hbar\frac{\partial \psi(\mathbf r,t)}{\partial t}=\left[-\frac{\hbar^2}{2m}
\nabla^2+V_{\rm ext}(\mathbf r)+g{\vert \psi(\mathbf r,t)
\vert}^2-\Omega L_z\right]
\psi(\mathbf r,t).\label{eq:GP}
\end{eqnarray}
Here the total external potential is given by 
$V_{\rm ext}(\mathbf r)=m\omega_{0}(x^2+y^2)/2+V_{\rm lattice}(\mathbf r)$,
$L_z=-i\hbar (x \frac{\partial}{\partial y}- y \frac{\partial}{\partial x})$ is the $z$ component of the anguler momenum operator, $g=4 \pi \hbar^2 a_{s}/m$ is the strength of
interaction with $a_{s}$ being the $s$-wave scattering length, and $ \gamma$ is the
phenomenolocical dissipative parameter \cite{SC,MO,EZ,TN,DS}.
The lattice potential created by blue-detuned laser beams arranged in the lattice
geometry is expressed as
\begin{eqnarray}
V_{\rm lattice}(\mathbf r) = \sum_{n_1,n_2}V_0\exp
\left[-\frac{\vert \mathbf r-  \mathbf r_{n_1,n_2}\vert^2}{(\sigma/2)^2}\right],
\label{eq:one}\end{eqnarray}
where $ \mathbf r_{n_{1},n_{2}}=n_{1}\mathbf a_{1}+n_{2}\mathbf a_{2}$.
We consider the triangular lattice geometry with the lattice constant $a$:
\ $\mathbf a_{1}=a(1,0)$, $ \mathbf a_{2}=a(-1/2,\sqrt{3}/2)$.
Throughout this paper, we scale the length and energy by
$a_{ho}=\sqrt{\hbar/m\omega_0}$ and
$\hbar\omega_0$.
We set the dimentionless interaction strength as $C=4 \pi a_{s}N/h_z=1000$, where
$N$ is the total particle number and $h_z$ is a height of the cylinder, and
the width of the laser beam as $ \sigma/a_{ho}=0.65$.
In our parameter set, the healing length is $ \xi/a_{ho}=0.12$.

\begin{figure}
\centerline{\includegraphics[height=2in]{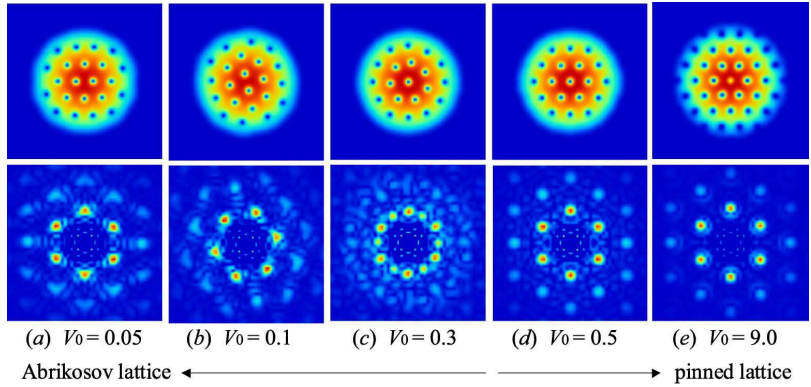}}
\caption{Density profiles (upper) and Structure factor profiles (lower)
of rotating condensates at $a/a_{ho}=2.2$, $\sigma/a_{ho}=0.65$ and $ \Omega/\omega_0=0.70$.
The figure show the vortex lattice structures corresponding to the intensity of the 
lattice potential
: $(a)$  $V_0=0.05$, $(b)$ $V_0=0.1$, $(c)$ $V_0=0.3$, $(d)$ $V_0=0.5$, $(e)$ $V_0=9.0$.}  
\label{fig:tau2}
\end{figure}

In our simulations, we first determine the ground-state condensate wavefunction
without rotation by setting $\Omega=0$ in the GP equation (\ref{eq:GP}).
Fig.~1(a) shows the condensate density profile in the lattice potential
without rotation.
We then dynamically evolve Eq.~(\ref{eq:GP}) with a fixed angular velocigy
$\Omega$ and look for equilibrium states.
In addition to the condensate density profile, we also look at the
density structure factor defined by
\begin{eqnarray}
S( \mathbf k)  = \int d \mathbf r n(\mathbf r) e^{-i \mathbf k \cdot \mathbf r},
\label{eq:Sk}
\end{eqnarray}
where $n(\mathbf r)=\vert \psi(\mathbf r)\vert^2 $ is condensate density.
The structure factor provides us information about the periodicity of the
condensate density.
For triangular lattice geometry, there are periodic peaks of regular
hexagonal geometry.
By looking at the position of peaks of the structure factor,
we can distinguish between the Abrikosov lattice and the pinned 
vortex lattice.

Fixing the lattice constant with $a/a_{ho}=2.2$, we investigated the transition 
of the vortex lattice structure by changing the intensity $V_0$ for various
angular velocities $ \Omega/\omega_0 $.
Fig.~2 shows the density profiles and the structure factors for 
$ \Omega/\omega_0=0.70$.
One can see that for weak lattice potentials 
($(a)$ $V_0=0.05$, $(b)$ $V_0=0.1$), vortices form the Abrikosov lattice,
which is incommensurate with the lattice potential.
By slightly increasing the intensity ($(c)$ $V_0=0.3$), vortices start to being
partially pinned by the lattice potentail.
For strong lattice potentials ($(d)$ $V_0=0.5$, $(e)$ $V_0=9.0$), all vortices 
are pinned by the lattice potential.
We thus observed the transition of the vortex lattice structure
from the Abrikosov lattice to the pinned lattice.

In order to quantify the structural phase transition of the vortex lattice
structure, we calculate the peak intensity of the structure factor $S({\bf K})$
at the lattice pinning point,
where \ $\mathbf K_{1}=2\pi/a(1,1/\sqrt{3})$, $ \mathbf K_{2}=2\pi/a(0,2/\sqrt{3})$.
In addition, we also calculate the lattice potential energy
\begin{eqnarray} E_{\rm lattice} = 
\int {d\mathbf r \psi^*(\mathbf r)V_{\rm lattice}(\mathbf r) \psi(\mathbf r)}.\label{eq:one}
\end{eqnarray}
%
In Fig.~3, we plot $S({\bf K})$ and $E_{\rm lattice}$ against of the
intensity $V_0$.
In Fig.~3(a) for $\Omega/\omega_0=0.70$, we find that $E_{\rm lattice}$ 
decreases gradually with vortices being partially
pinned, and reaches constant when all vortices are pinned for $V_0>0.4$.
Correspondingly $S({\bf K})$ increases gradually and finally becomes constant
when all vortices are pinned.
From these results together with directly looking at the condensate 
density profile, we conclude that the structural phase transition
occurs at $V_0=0.4$.
\begin{figure}
\begin{minipage}{0.5\linewidth}
\centering
\centerline{\includegraphics[height=1.8in]{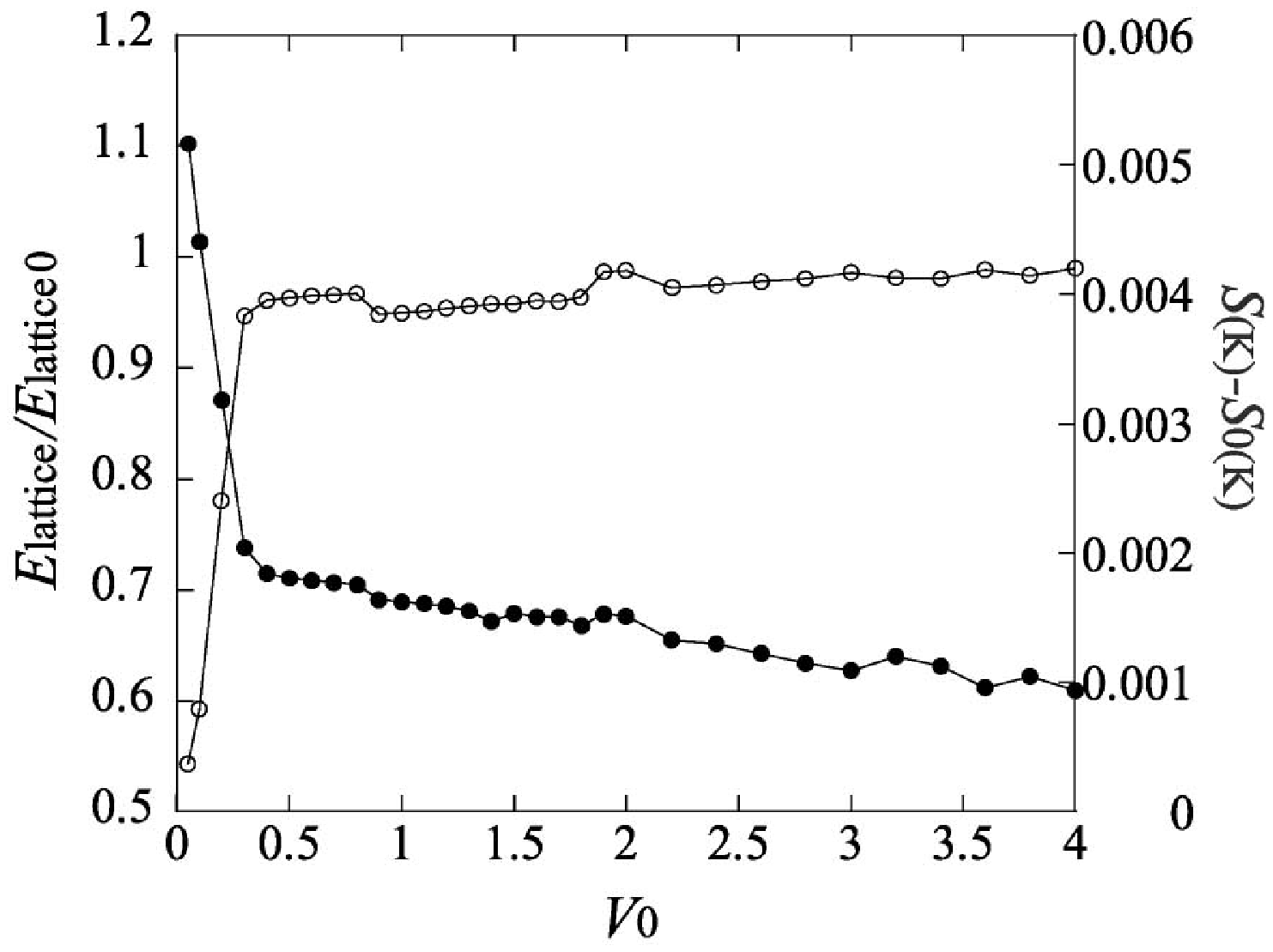}} {$(a)$}
\end{minipage}
\begin{minipage}{0.5\linewidth}
\centering
\centerline{\includegraphics[height=1.8in]{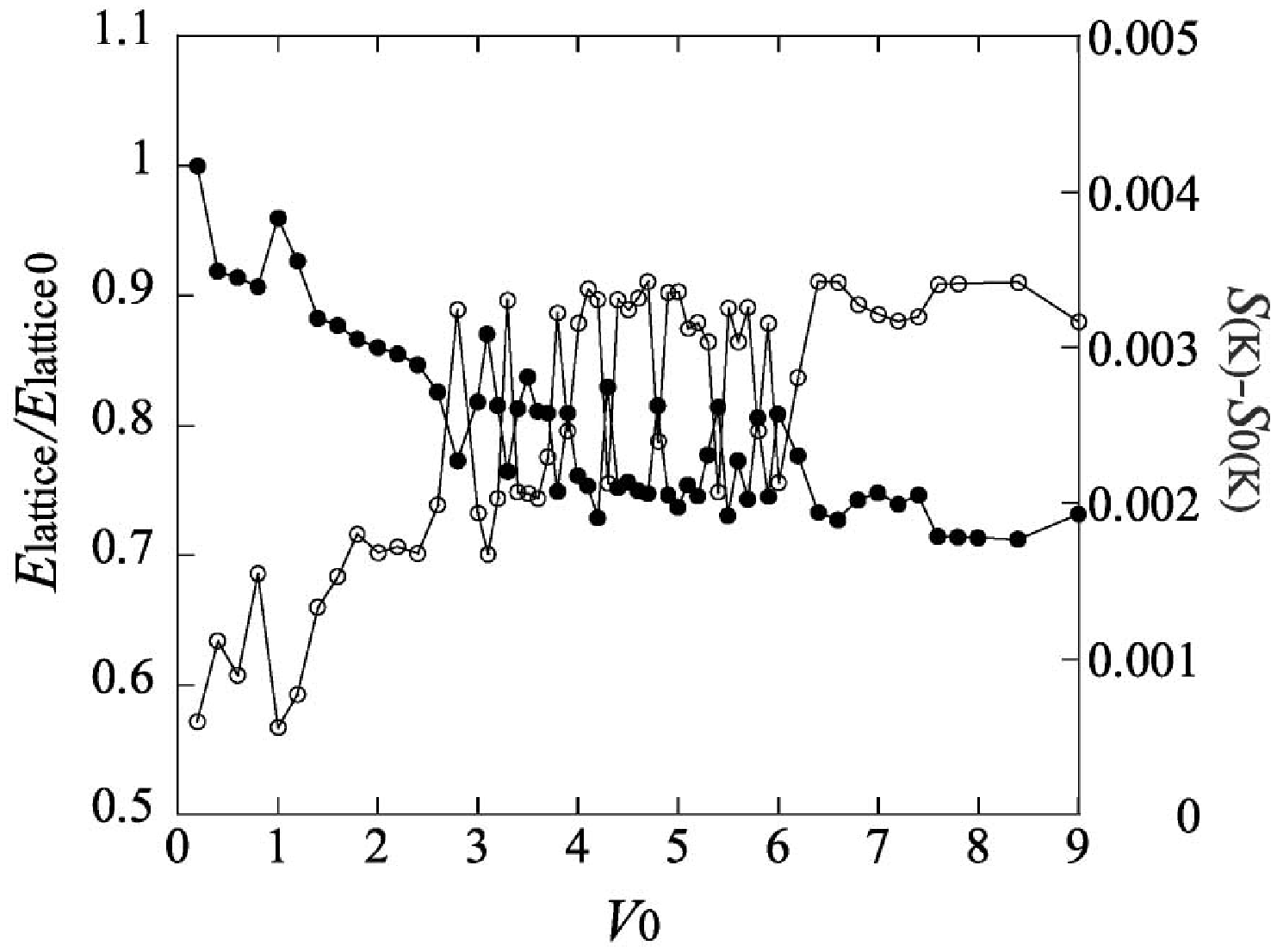}}{$(b)$}
\end{minipage}
\caption{Lattice potential energy ($ \bullet $) and peak intensity of the structure factor
at the lattice pinning point ($ \circ $) with $a/a_{ho}=2.2$, 
$\sigma/a_{ho}=0.65$ corresponding to the angular velocity:
$(a)$ $\Omega/\omega_0=0.70$, $(b)$ $\Omega/\omega_0=0.55$.
Here $E_{\rm lattice0}$ and $S_0({\bf K})$ is the lattice energy and the peak intensity of the structure factor
at the lattice pinning point of ground state for the intensity respectivity.}  
\label{fig:tau2}
\end{figure}
Fig.~3(b) shows the results for $ \Omega/\omega_0=0.55$.
In this case, there is an intermediate domain where the Abrikosov lattice 
and the pinnd lattice coexist.
In this domain, the vortex lattice structure cannot be categorically
determined because of the competition between the vortex-vortex interaction 
and the lattice pinning effect.

In Fig.~4, we map out the phase diagrams of the vortex lattice structures against $ \Omega/\omega_0$ and $ V_0$.
The lower domain is the Abrikosov lattice domain, while the upper domain is the 
pinned lattice domain.
The intermediate domain represents coexisting state of the Abrikosov lattice 
and the pinnd lattice.
From Fig.~4(a) for $a/a_{ho}=2.2$, 
we find that the pinning intensity takes the minimum value 
$V_{\rm min}=0.4$ at $ \Omega/\omega_0=0.70$.
Similarly, we find 
$V_{\rm min}=0.7$ at $\Omega/\omega_0=0.83$ for
$a/a_{ho}=2.0$ (as shown in Fig.~4(b)), and 
$V_{\rm min}=1.0$ at $\Omega/\omega_0=0.65$ for
$a/a_{ho}=2.4$ (as shown in Fig.~4(c)).
\begin{figure} [h]
\begin{minipage}{0.5\linewidth}
\centering
\centerline{\includegraphics[height=1.79in]{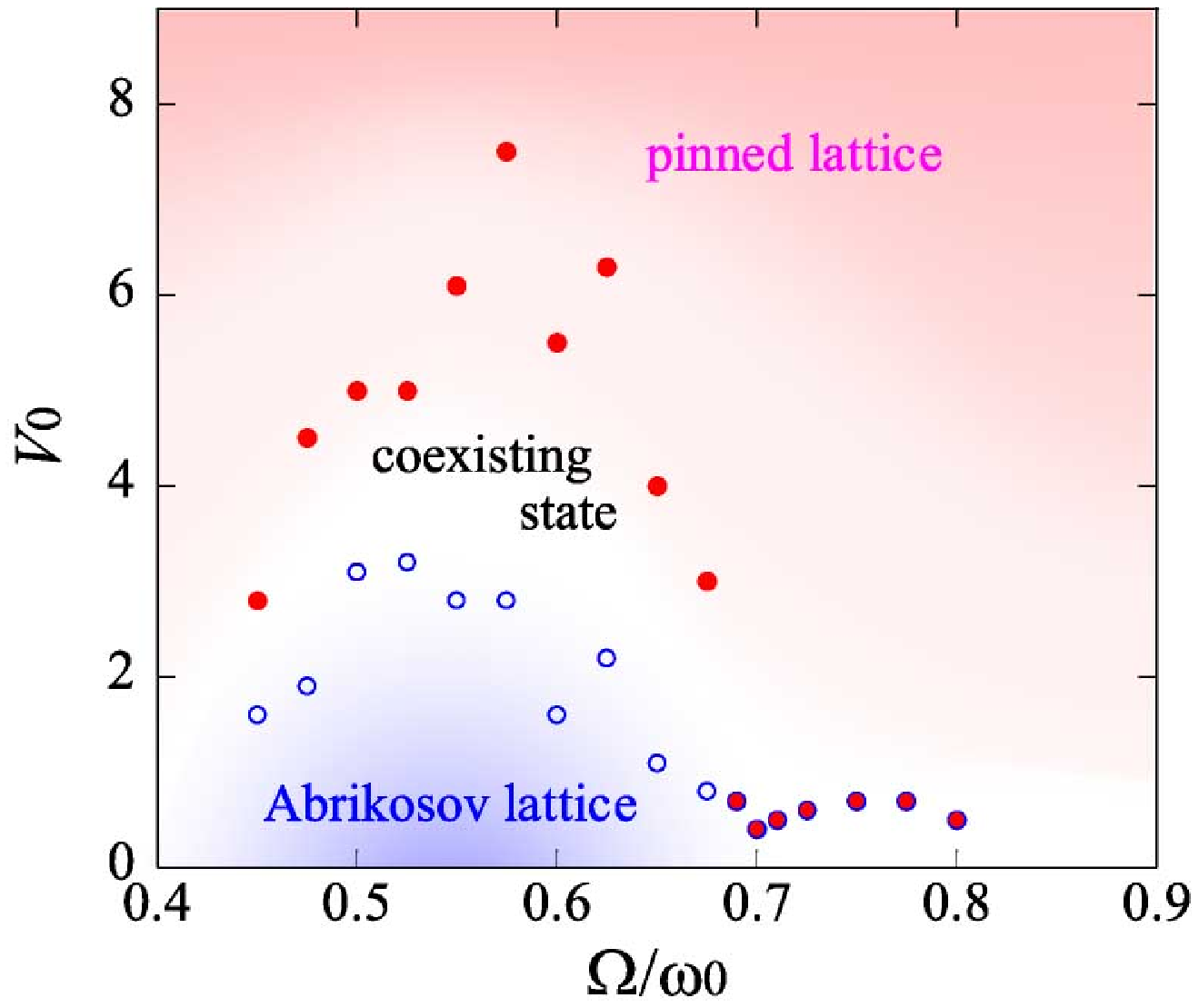}}{$(a)$}
\end{minipage}
\begin{minipage}{0.5\linewidth}
\centering
\centerline{\includegraphics[height=1.79in]{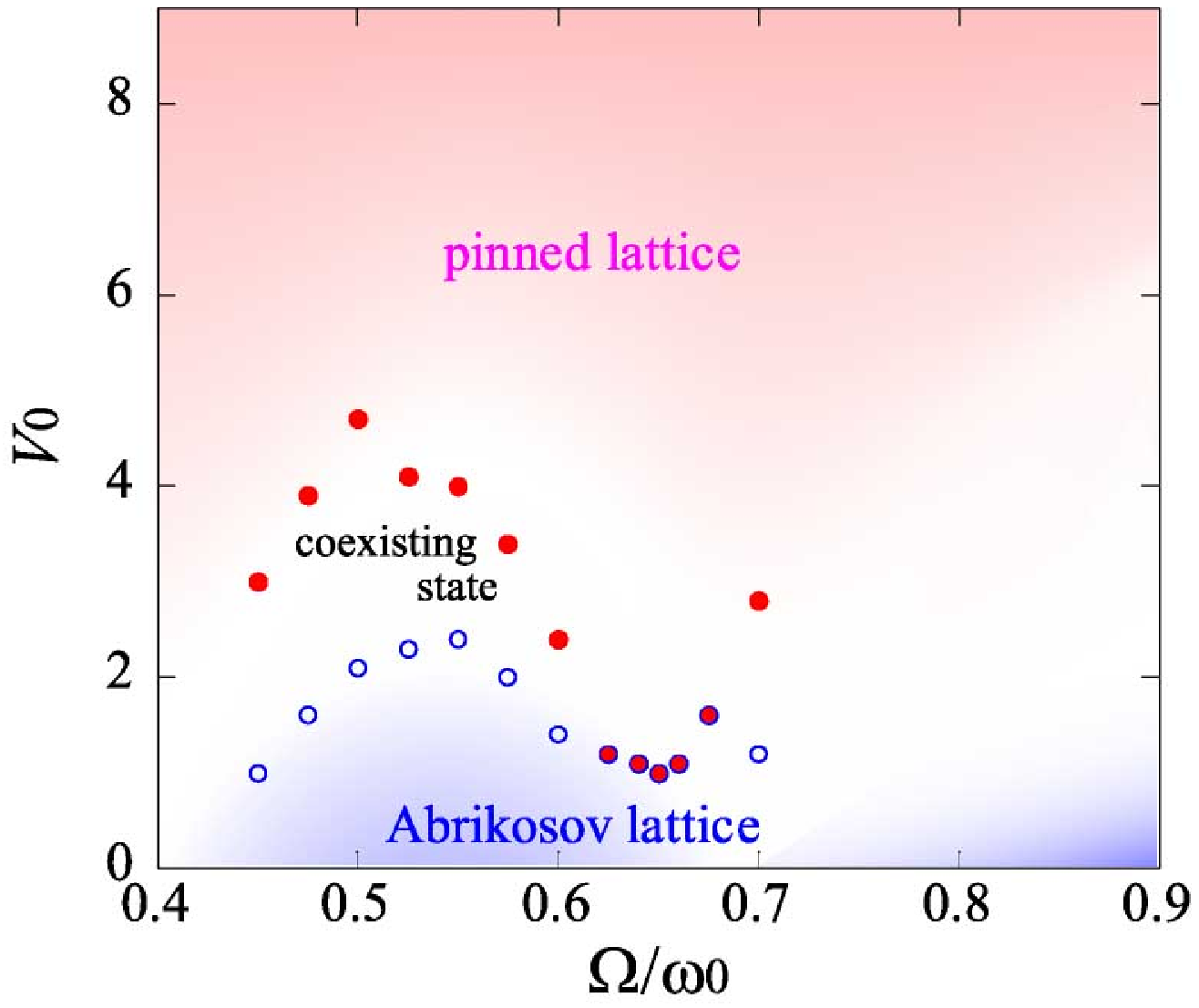}}{$(b)$}
\end{minipage}
\begin{minipage}{0.5\linewidth}
\centering
\centerline{\includegraphics[height=1.79in]{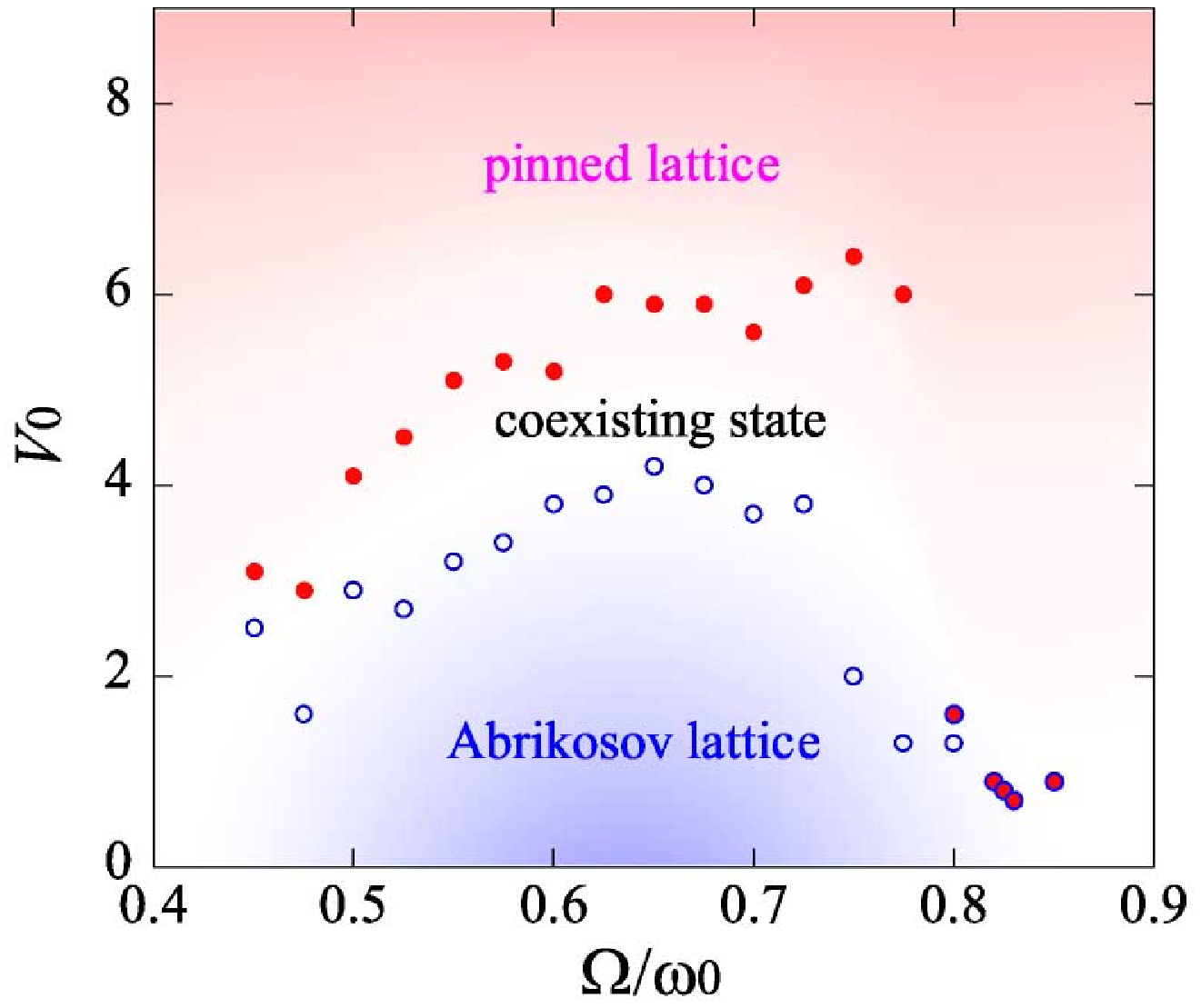}}{$(c)$}
\end{minipage}
\begin{minipage}{0.5\linewidth}
\centering
\centerline{\includegraphics[height=1.79in]{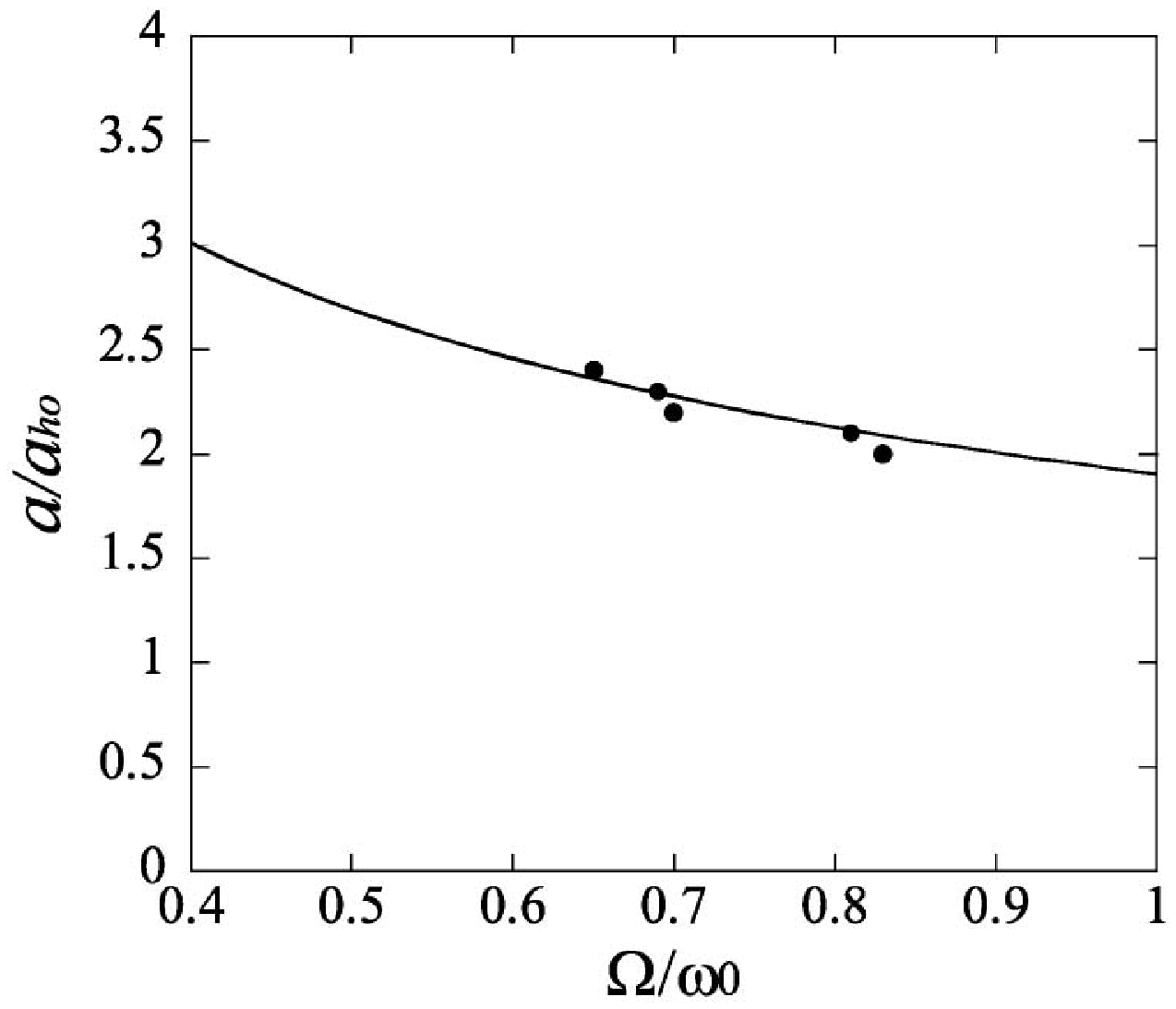}}
{$(d)$}
\end{minipage}
\caption{(1) Phase diagrams  for the vortex lattice structures:
$(a)$ $a/a_{ho}=2.2$, $(b)$ $a/a_{ho}=2.4$, $(c)$ $a/a_{ho}=2.0$.
(2) The angular verocity corresponding the minimum pinning intensity:
{$(d)$}.
The solid line represents Eq. (5).}  
\label{fig:tau2}
\end{figure}

The minimum values of the pinning lattice intensity $V_0$ for each 
lattice constant can be understood as follows.
When vortices form the trianglular lattice and undergo rigid rotation with angular
velocity $ \Omega $, the lattice constant is expressed as a function of angular
velocity by
\begin{eqnarray}
a_v(\Omega)/a_{ho}=\sqrt{\frac{2}{\sqrt{3}} \frac{\kappa}{2\Omega/\omega_0}},
\label{eq:a_omega}
\end{eqnarray}
where $ \kappa=h/m $ is the circulation of a singly-quantized vortex.
In Fig.~4(d), we plot the lattice constant $a/a_{ho}$ against $ \Omega/\omega_0 $ giving the
minimum pinning intensity.
We find that it is well fitted by the funcion in Eq.~(\ref{eq:a_omega}). 
This means that the pinning intensity $V_0$ takes minimum value when $a_v(\Omega)$
matches to the lattice constant $a$. 
When this ``matching relation'' is satisfied, weak lattice potential has
a stronger pinning effect than vortex-vortex interaction, which leads to
a sharp transition of the vortex lattice structure. 

In summary, we have studied the transition of the vortex lattice structures of 
Bose-Einstein condensates in a rotating triangular lattice potential.
We showned that the transition is determined depending on parameters of the lattice potential.
We also found that the lattice potential has a strong pinning effect when the matching relation
is satisfied.
For a future study, we will discuss the transition of vortex lattice structures for various lattice potential geometries. 

We thank S. Konabe and S. Watabe for useful discussions and comments. 
We also thank N. Sasa for helpful comments on the numerical simulations.

\end{document}